\documentclass[11pt,letterpaper]{article}

\usepackage{authblk}

\usepackage{amsmath, amsthm, amssymb,color,fullpage,url,booktabs}  
\usepackage[pdftex]{hyperref} 
\usepackage{graphicx}

\newcommand{\nc}{\newcommand}
\nc{\rnc}{\renewcommand}

\newcommand{\bra}[1]{\left\langle #1\right|}
\newcommand{\ket}[1]{\left|#1\right\rangle}

\DeclareMathOperator{\tr}{tr}
\DeclareMathOperator{\rank}{rank}

\def\be#1\ee{\begin{equation}#1\end{equation}}
\def\bea#1\eea{\begin{eqnarray}#1\end{eqnarray}}
\def\beas#1\eeas{\begin{eqnarray*}#1\end{eqnarray*}}
\def\ba#1\ea{\begin{align}#1\end{align}}
\def\bas#1\eas{\begin{align*}#1\end{align*}}
\def\bpm#1\epm{\begin{pmatrix}#1\end{pmatrix}}

\newtheorem{thm}{Theorem}
\newtheorem*{thm*}{Theorem}

\newtheorem{dfn}[thm]{Definition}
\newtheorem{proto}{Protocol}

\makeatletter
\newtheorem*{rep@theorem}{\rep@title}
\newcommand{\newreptheorem}[2]{%
\newenvironment{rep#1}[1]{%
 \def\rep@title{#2 \ref{##1} (restatement)}%
 \begin{rep@theorem}}%
 {\end{rep@theorem}}}
\makeatother

\newreptheorem{thm}{Theorem}
\newreptheorem{lem}{Lemma}

\def\benum{\begin{enumerate}}
\def\eenum{\end{enumerate}}
\def\bit{\begin{itemize}}
\def\eit{\end{itemize}}

\nc{\todo}[1]{\textcolor{red}{todo: #1}}

\def\begsub#1#2\endsub{\begin{subequations}\label{eq:#1}\begin{align}#2\end{align}\end{subequations}}
\nc\qand{\qquad\text{and}\qquad}
\nc\mnb[1]{\medskip\noindent{\bf #1}}

\usepackage{boxedminipage}

\usepackage{amsmath}
\usepackage{amssymb}

\usepackage[T1]{fontenc}
\usepackage[utf8]{inputenc}
\usepackage{authblk}

\usepackage{enumerate}
\usepackage{enumitem}
\usepackage[]{units}
\usepackage{amsfonts}
\usepackage{amsmath}
\usepackage{mathtools}
\usepackage{amsthm}
\usepackage{bm}
\usepackage{mciteplus}

\DeclareMathOperator{\Tr}{Tr}

\newcommand\I{\mathcal{I}}

\newcommand\Psym{\mathcal{P}}
\newcommand\F{\mathcal{F}}
\newcommand\A{\mathcal{A}}
\newcommand\C{\mathcal{C}}
\newcommand\Chi{\chi}

\title{A Classical Model Correspondence for $G$-symmetric Random Tensor Networks}

\begin{document}

\author{Erica Morgan}

\author{Fernando G.S.L. Brand\~ao}

\affil{Institute of Quantum Information and Matter, California Institute of Technology, Pasadena, CA}

\maketitle

\begin{abstract}
We consider the scaling of entanglement entropy in random Projected Entangled Pairs States (PEPS) with an internal symmetry given by a finite group $G$. We systematically demonstrate a correspondence between this entanglement entropy and the difference of free energies of a classical Ising model with an addition non-local term. This non-local term counts the number of domain walls in a particular configuration of the classical spin model. We then make use of this correspondence to argue for an area law scaling with well-defined topological entanglement entropy when the bond dimensions are sufficiently large. The topological entanglement entropy is shown to be $\log{|G|}$ for a simply connected region and results from a difference in the number of domain walls of ground state energies for the two spin models. 
\end{abstract}


\section{Introduction}

For gapped models of local Hamiltonians on a finite dimensional lattice, the entanglement between a region $A$ and the rest of the system is expected to be given by \cite{Q1,Eisert1}:
\begin{equation} \label{arealawstrong}
S(A) = \alpha |\partial A| - \gamma + o(1).
\end{equation}
Here $S(A) := - \tr(\rho_A \log \rho_A)$ is the von Neumann entropy of the reduced density matrix $\rho_A$ of the groundstate in region $A$, $\partial A$ is the boundary size of $A$, and $\alpha$, $\gamma \geq 0$ constants, both of which are independent of the choice of the region. The term $o(1)$ stands for subleading corrections which go to zero with growing surface area of the region $A$. 

The first term in Eq. (\ref{arealawstrong}) shows that entanglement of the region scales with its boundary, instead of its volume as is the case for the majority of quantum states. It shows that the entanglement inherits the locality of the interactions, though we note that this is not always the case if the system is gapless. The second term $\gamma$ is an additive correction to the area law scaling and is a signature of topological order in the system \cite{Q3,levin2006detecting}. It is known as the topological entanglement entropy. For this quantity to be well defined, it is imperative that $\alpha$ is independent of the choice of the region. 


However, Eq. (\ref{arealawstrong}) does not always hold. For example, we cannot have $\alpha$ independent of $A$ for systems with disorder. For systems with translation invariance, it is possible that Eq. (\ref{arealawstrong}) applies generally. Even the problem of proving that $S(A) \leq O(|\partial A|)$ is a major open question in dimensions larger than one (see \cite{Hastings1, arad2013area, brandao2013area} for the 1D case). Therefore, it is an interesting question to identify families of models for which one can argue Eq. (\ref{arealawstrong}) gives the right scaling of entanglement. One particular example are renormalization fixed-point models, such as string-net models \cite{flammia2009topological,kitaev2006anyons,levin2005string}. There Eq. (\ref{arealawstrong}) holds with no correction term (denoted by $o(1)$ there). However, it remains a challenge to go beyond such models. We note that numerically one can nonetheless verify the right form in many cases for the 2-Renyi entropy.
 
Given the difficulty of finding a general argument for the scaling of Eq. (\ref{arealawstrong}), one can ask whether it holds in the generic case, i.e. for most states in the class considered. If we consider ground states of gapped Hamiltonians as the class of states, this too seems like a very hard problem. Another approach is to consider states which automatically satisfy the weaker form of area law $S(A) \leq O(|\partial A|)$ and ask whether generically the entanglement scales as Eq. (\ref{arealawstrong}). This is the path we take in this paper. In particular, we consider Projected Entangled Pairs States (PEPS), a useful class of tensor networks states with entanglement structure mimicking the locality of the interactions \cite{verstraete2004renormalization}. They are known to well approximate thermal states of local models in any dimension \cite{hastings2006solving, molnar2015approximating}, and are expected to also approximate well the groundstates for gapped models (yet a proof for the latter is challenging). Seeing how the topological entanglement entropy manifests in such models may provide a clue towards proving the same in more general settings.

The case of random PEPS where each tensor is completely random was considered in an insightful paper by Hayden \textit{et al} \cite{Hayden1}. Their motivation was to study features of the AdS/CFT correspondence in a tensor network picture. Among other findings, they showed that the entanglement entropy scales as
 \begin{equation} \label{arealawstrongHayden}
S(A) = \log(D) |\partial A_{RT}| + o(1).
\end{equation}
where $D$ is the so-called bond dimension of the PEPS (the dimension of internal degrees of freedom of the state responsible for its entanglement) and $|\partial A_{RT}|$ is the length of the Ryu-Takayanagi surface in the bulk for a boundary region $A$. The error term $o(1)$ goes to zero as the bond dimension grows. The authors demonstrated a mapping from the problem of computing the entanglement entropy to the problem of computing the difference of free energies due to a domain wall in a certain classical Ising model associated to the state (where the $\log(D)$ becomes the inverse temperature of the model). We choose to extend this construction to the setting of a PEPS describing a 2D physical region, e.g. on a square lattice with one physical index on each tensor of the network, but nonetheless note the relevance of our results and studying random tensor networks in general when considering holography. In particular, one might be interested in constructions like ours when attempting to understand holographic coherent states \cite{qi2017holortn1}.

In order to extend this construction into our setting of interest, we must acknowledge two major drawbacks of the original approach. The first is that the result relies on the "physical" boundary dimension growing arbitrarily large. In the condensed matter setting, we care about systems of particles with finite degrees of freedom (e.g. qubits with physical dimension 2). We address this later in the paper and note that a rigorous account of our results in this setting remains elusive. Another drawback is that the topological entanglement entropy $\gamma$ is always zero in the generic case of fully random PEPS. This is expected, as one expects to be able to adiabatically connect them to a trivial state. We work around this drawback by providing an additional constraint in the form of an internal symmetry on each tensor. 

In more detail, in Ref. \cite{Schuch1}, Schuch, Cirac and Perez-Garcia gave a construction of PEPS with topological order associated to a finite group--incidentally a quantum double model--by imposing that the virtual degrees of freedom are symmetric under a regular unitary representation of the group. They also showed that for renormalization fixed-point PEPS, Eq. (\ref{arealawstrong}) holds with $\gamma = \log |G|$, with $|G|$ the order of the group associated to the topological order (e.g. $G = Z_2$ for the toric code). 
 
In this paper we take an approach inspired by \cite{Hayden1} and using the internal symmetry described by \cite{Schuch1} in an effort to calculate the form of entanglement entropy in random PEPS states with an internal symmetry associated to a finite group. We argue that generically,
\begin{equation} \label{maininequality}
S(A) = \log(D) |\partial A| -  \log|G| + o(1),
\end{equation}
for bond dimension $D$ scaling with the volume of the region $A$. We discuss briefly conditions under which this may be true when the phyiscal bond dimension, $d$, is kept small.
	
In analogy to Ref. \cite{Hayden1}, we also obtain our result by mapping the problem to computing the free energy of a classical model. Interestingly, in the case of non-trivial topological order, the model has (in addition to the usual local terms) a non-local energy term which counts the number of domain walls of a configuration. This is the term which gives the $\log|G|$ correction to entanglement at high bond dimension or sufficiently low temperature.

The organization of the rest of the paper is as follows. In section \ref{sec:3} we derive a classical model associated to a random tensor network with a local virtual symmetry, arriving at an Ising action with an additional non-local term. We then use this to compute the entropy for a system with local symmetry. We finish the section by briefly addressing the constraint of small physical dimension. Finally, in section \ref{sec:8} we summarize our results and provide some comments and connections back to the holography setting as well as future directions. Appendix A generalizes our construction to higher order R\'{e}nyi entropies.



\section{Computing Entanglement Entropy with G-Symmetry} \label{sec:3}
	Our aim is to arrive at an expression for the Von Neumann entropy for a $G$-injective random tensor network by bounding it below by the second R\'{e}nyi entropy and above by the Schmidt rank. In this section, we first work through computing the R\'{e}nyi entropy via a difference of free energies of a particular Ising model.
	
	To start, we consider a two-dimensional square lattice throughout the entirety of this work for simplicity, though these results straightforwardly generalize to any lattice where the correspondence is then to a classical model on the associated interaction graph. The sites of the lattice are indexed by $x$. We associate a finite dimensional vector space 

\begin{equation}
    V_x := \bigotimes_{x \sim y} V_{x, v} \otimes V_{x, \partial}  \cong   (\mathbb{C}^{D_{xy}})^{\otimes 4} \otimes \mathbb{C}^{d_x}
\end{equation}
    to each site. The tensor product runs over the neighbor sites $y$ of each vertex $x$. To each neighbor we associate a $D_{xy}$-dimensional vector space. Each site also has a $d_x$-dimensional vector space $V_{x, \partial}$, which will be associated with the physical degrees of freedom residing at each site of the lattice. Let $D_x$ be the total dimension of the space $V_x$. We will use different notation than \cite{Hayden1}, and refer to the 'physical' indices rather than 'dangling' indices to emphasize the setting.

\begin{figure}
    \centering
    \includegraphics{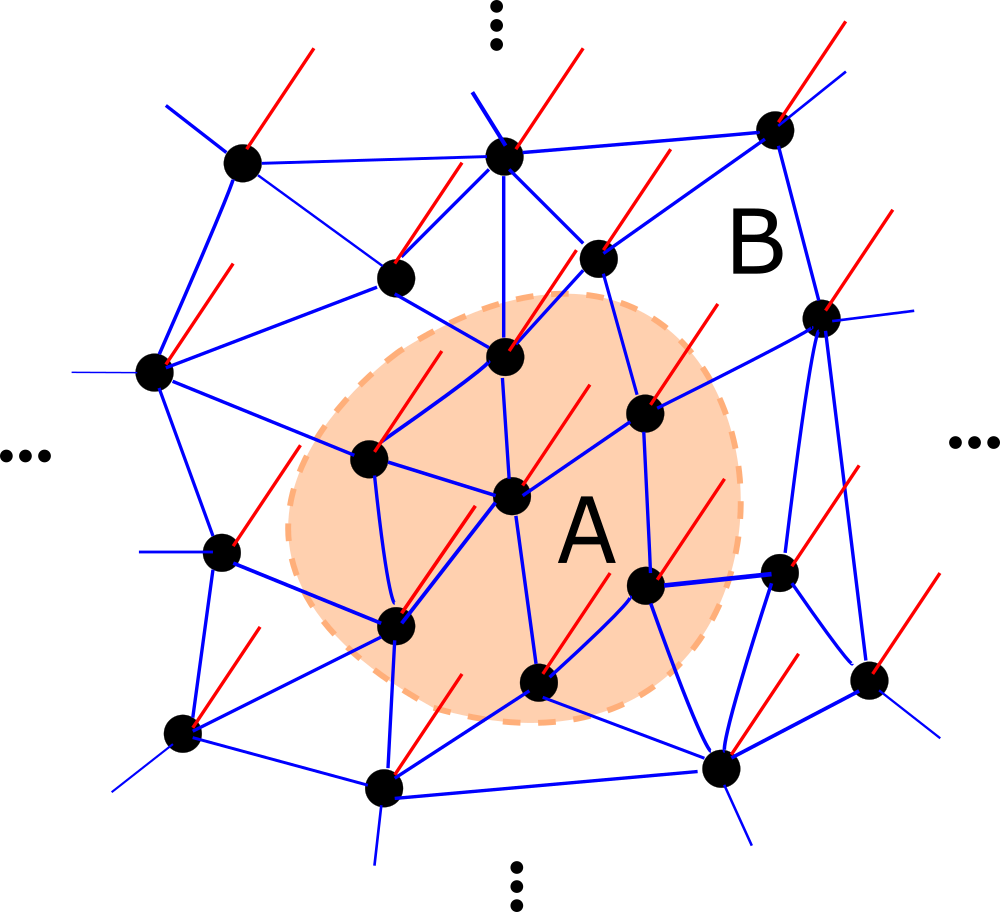}
    \caption{Example of an arbitrary PEPS tensor network divided into two regions. The blue legs are associated with the virtual Hilbert spaces on which reside maximally entangled states. The red legs are the boundary dangling legs. Note here that there are 7 indices in $A$ and $|\partial A| = 13$.}
    \label{fig:1}
\end{figure}

    Let $\ket{V_x}$ be a random state on $V_x$, chosen from the Haar measure. Let $\ket{xy}$ be a maximally entangled state on $V_{y, v} \otimes V_{x, v}$, for two neighboring sites $x, y$. We start by considering a random (non-normalized) PEPS state defined as (see Figure \ref{fig:1} for an example):
\begin{equation}
	\ket{\Psi} = \left( \bigotimes_{ x \sim y   } \bra{xy} \right) \left( \bigotimes_{x} \ket{V_x} \right),
\end{equation}
    with the first tensor product running over all maximally entangled states acting on the vectors spaces $V_{y, v} \otimes V_{x, v}$ for all neighboring $x, y$, and the second over all sites. This leads to a generic form of the (non-normalized) density operator given by
\begin{equation} \label{eq:6}
	\rho = \Tr_P{\left(\bigotimes_{x \sim y } \ket{xy}\bra{xy} \prod_{x} \ket{V_x}\bra{V_x}\right)}.
\end{equation}
    The subscript $P$ indicates a trace over the vector spaces $V_{x, v}$, corresponding to the contraction of internal indices. Note that $\rho$ is a linear function of independent pure states living on the vertices. 
    
\begin{figure}
\centering
\includegraphics[scale=0.5]{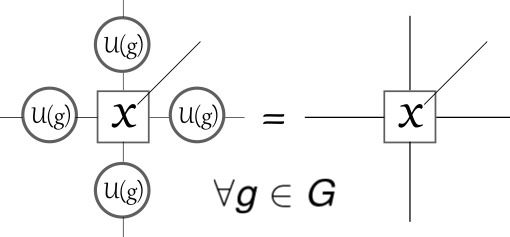}
\caption{Graphical representation of the symmetry constraint on a given vertex with four associated edges.}
\label{fig:2}
\end{figure}

	We also wish to constrain each vertex in the lattice so that the virtual indices remain invariant under operation by a unitary representation of a finite group $G$. We denote such representation by $U(g)$ and further require that it be a regular representation for reasons that will become apparent later in this section. It is worth noting that these representations must be such that the dimension of the representation or the total dimension of a direct sum of such representations equals the bond dimension. The latter case may by of further interest, such as in describing bulk gauge fields \cite{qi2017holortn1}, though we do not consider it here. In any case, we will consider a tensor network--still residing on a square lattice--made from states $\ket{\psi}_x$ on $V_x$ such that 
	
\begin{equation} \label{eq:29}
	U(g)^{\otimes 4} \otimes \I \ket{\psi_x} = \ket{\psi_x}, \hspace{0.2 cm} \forall g \in G
\end{equation}
	where the unitary representations, $U(g)$, act upon the 4 virtual subspaces surrounding a particular vertex, the identity is acting upon the physical subspace, and $g \in G$ are group elements. We describe this visually in Figure \ref{fig:2}. The more general case follows where we choose the symmetry such that a unitary representation lies on every virtual edge associated to a vertex. We might also further generalize this construction by allowing for different weights associated with each representation. 

    Nonetheless, our primary objective with this construction is to compute the second R\'{e}nyi entropy for a region $A$:
\begin{equation} \label{eq:8}
	S_2(A) = - \log{ \frac{\Tr{\rho_A^2}}{(\Tr{\rho_A})^2} }.
\end{equation}
where $\rho_A$ is the reduced density matrix of a region $A$. 

\subsection*{Ising Model Correspondence for Topologically Nontrivial PEPS}
    Rather than starting from the entropy, we will begin by rigorously establishing a correspondence between the averaged trace over the choice of a particular PEPS tensor network and the partition function of a classical Ising model. We will then return to the entropy further below and use this correspondence to sketch a computation thereof. The correspondence we establish parallels that of ref. \cite{Hayden1} at the start and departs into further detail and an inclusion of the non-trivial virtual symmetry. This result will form the central component of the paper which allows us to make an evidenced claim to the form of the entropy in the non-trivial case.

\begin{thm}[Projected Two-copy Trace/Modified Ising Model Correspondence]
	Let $Z_1 = \Tr{\rho_A^2}$ for a reduced density matrix on a region $A$ which is described by a PEPS with local invariance given by a finite group, $G$ (see Eq. \ref{eq:29}). Then the Haar average of $Z_1$ equals the partition function of a modified classical Ising model of spins $\{ s_x \}$  residing on the same lattice with Ising action given by
\begin{equation}
	\A[\{s_x\}] = - \sum_x \frac{1}{2}\log{d_x}(s_x h_x-1) - \sum_{x \sim y} \frac{1}{2}\log{D_{xy}}(s_x s_y-1) - \eta \log{|G|} + \text{const},
\end{equation}
	where $d_x$ is the physical bond dimension of the tensor at $x$, $D_{xy}$ the virtual space bond dimension between tensors at $x$ and $y$, and $D_x$ the total dimension of the tensor at $x$. The variable $h_x$ is the boundary field which is 1 for $x \in \overline{A}$ and -1 for $x \in A$. The non-local term is characterized by the number of domain walls in a particular spin configuration, $\eta$, and the cardinality of the finite group $G$.

	If we instead have an average over $Z_0 = (\Tr{\rho_A})^2$, then we have the same correspondence but with a uniform boundary pinning field, i.e. $h_x = 1, \forall x$. 
\end{thm}

\begin{proof}
	We begin by considering $Z_1$ (and $Z_0$) in a slightly more convenient form,
\begin{equation}
	Z_1 = \Tr{\rho_A^2} = \Tr{[(\rho \otimes \rho) \F_A]},
\end{equation}
\begin{equation}
	Z_0 = \Tr{\rho_A}^2 = \Tr{[\rho \otimes \rho]}.
\end{equation}
	where the second equality for $Z_1$ is a result of using the so-called 'swap trick'; the operator $\F_A$ swaps the two copies of the system in region $A$ only. It is important to note for late that $\F_A$ acts only on the physical degrees of freedom at each vertex and not the internal Hilbert spaces. In order to constrain our quantum state to one where local tensors are $G$-symmetric, we restrict otherwise random tensors to symmetric subspaces. To do this, we need to introduce our symmetric subspace projector,
\begin{equation}
	\Psym_{x} \equiv \frac{1}{|G|} \sum_{g_x \in G} U(g_x)^{\otimes \nu_x} \otimes \I.
\end{equation}

	We can then project the otherwise Haar random states living on each vertex onto the symmetric subspace using this projector to obtain a slightly modified expression for the numerator of the entropy, $Z_1$, from the trivial case.  We will do this on a uniform square lattice ($n_x = 4, \forall x$) for simplicity. The formalism extends to any random PEPS tensor network with local $G$-symmetry in terms of operators acting upon virtual subspaces for each edge in the interaction graph. 
	
    By inserting our subspace projector and eq. (\ref{eq:6}) into our expression for $Z_1$ and then taking the Haar average we obtain,
\begin{equation}
	Z_1 = \Tr{\left[ \bigotimes_{x \sim y} \ket{xy}\bra{xy}^{\otimes 2} \F_A \prod_{x} \Psym_{x}^{\otimes 2} \overline{\ket{V_x}\bra{V_x}^{\otimes 2}} \right]}.
\end{equation}

	Note that the average of $Z_1$ follows through to just an average over the random vertex states $\ket{V_x}$ since the trace is a linear operator and the rest of the expression is not random. Additionally, we do not have the adjoint of the projector on the right side of the average as it commutes through the average. 
	
	To compute this average, we take an arbitrary reference state $\ket{0_x}$ so that our random states are just random unitary rotations from the reference, $\ket{V_x} = U_x \ket{0_x}$. We can then take the average by integration of unitaries with respect to the Haar measure \cite{Harrow1}, 
\begin{equation}
	\overline{\ket{V_x}\bra{V_x} \otimes \ket{V_x}\bra{V_x}} = \int dU_x (U_x \otimes U_x)(\ket{0_x}\bra{0_x} \otimes \ket{0_x}\bra{0_x}) (U_x^{\dagger} \otimes U_x^{\dagger}) = \frac{I_x+\F_x}{D_x^2+D_x}.
\end{equation}
	In this expression, $I_x$ denotes an identity operator and $\F_x$ the swap operator each acting on the two copies of the Hilbert space at vertex $x$. $\F_x$ is defined in a similar way as $\F_A$, swapping the two copies of the state at vertex $x$, including the virtual spaces unlike $\F_A$. Replacing the average in the expression with this result, we now have an expression consisting of $2^N$ terms for $N$ vertices. Each of the terms indicates all combinations of either an identity or swap on each vertex,
	
\begin{equation}
	\overline{Z_1} = \frac{1}{\prod_{x} (D_x^2+D_x)} \Tr{\left[ \bigotimes_{x \sim y} \ket{xy}\bra{xy}^{\otimes 2} \F_A \prod_{x} \Psym_{x}^{\otimes 2} (\I_x+\F_x) \right]}.
\end{equation}	
	
	In order to simplify this expression, we introduce a classical spin variable, $s_x$, where $s_x = +1$ to indicate the use of an identity and $s_x = -1$ to indicate the use of a swap operator. This representation forms the basis for our correspondence and leads us to consider $\overline{Z_1}$ as a partition function of spins $\{s_x\}$ such that
\begin{equation} \label{eq:9}
	\overline{Z_1} = \sum_{\{s_x\}} e^{-\A [\{s_x\}]},
\end{equation}
	
	and
\begin{equation} \label{eq:17}
	e^{-\A[\{s_x\}]} \equiv \frac{1}{\prod_x (D_x^2+D_x)} \Tr{\left[ \bigotimes_{x \sim y} \ket{xy}\bra{xy}^{\otimes 2} \F_A \prod_x \Psym_{x}^{\otimes 2} \prod_{x\text{ s.t. }s_x=-1} \F_x \right]}.
\end{equation}
	
	The trace in this equation is over an expression which is a tensor product of terms acting on different parts of the Hilbert spaces describing two copies of an $N$-vertex PEPS. The first part of the expression inside the trace consists of all maximally entangled density states describing internal lines in the network and acts trivially on the dangling indices. The swap $\F_A$ acts only on the physical indices of the region $A$ and the symmetric subspace projects acts non-trivially on only the virtual indices. The remainder of the expression is either a swap or identity depending on the spin variable. Each of these swaps act on physical and virtual spaces independently and thus can be considered as a tensor product of swaps on each space.
	    
	Altogether, then, we are motivated to split the trace as the expression within is a tensor product of operators acting independently on the physical and virtual Hilbert spaces. We denote the physical space by the subscript $\partial$ and virtual space by the subscript $v$ such that,
\begin{multline}
	e^{-\A[\{s_x\}]} = \frac{1}{\prod_x(D_x^2+D_x)} \Tr_{\partial}{\left[\F_A \prod_{x\text{ s.t. }s_x=-1} \F_{x,\partial}\right]} \\
	\times \Tr_{v}{\left[\bigotimes_{x \sim y} \ket{xy}\bra{xy}^{\otimes 2} \prod_x \Psym_x^{\otimes 2} \prod_{x\text{ s.t. }s_x=-1}\F_{x,v}^{\otimes 4}\right]}.
\end{multline}
        
	The trace over physical indices can be simplified by considering one operator which is either a swap or identity depending upon the combination of swaps in the expression. In order to do this, we introduce a variable $h_x$ (the same boundary field as in \cite{Hayden1}),
\begin{equation}
h_x = 
\begin{cases}
	+1, & x \in \bar{A} \\ 
	-1, & x \in A.
\end{cases}
\end{equation}
	The trace over the physical Hilbert space is then over an operator which is a product of swaps depending whether $x \in A$ or not and which term in the partition function we are concerned with. This allows us to write the trace as 
\begin{equation}
	\Tr_{\partial} {\left[ \F_A \prod_{x\text{ s.t. }s_x=-1} \F_{x,\partial} \right]} = \Tr_{\partial} {\left[ \prod_x \F_{x,\partial}^{\frac{1}{2}(s_x h_x +3)} \right] }.
\end{equation}
	We use this particular expression to describe an operator which is a swap or identity as desired for a particular combination of $s_x$ and $h_x$. The result is a trace over a product of operators acting independently on the physical Hilbert spaces associated with each vertex, $x$. 
	
	As such, the trace amounts to a product of traces on each space individually. Each of these traces is $d_x^2$ if it is over an identity and $d_x$ if it is over a swap, where $d_x$ is the dimension of the physical Hilbert space of the vertex $x$. This depends on $s_x$ and $h_x$ in the same way the operator does so that our trace above is simply
\begin{equation} \label{eq:14}
	\Tr_{\partial} {\left[ \prod_x \F_{x,\partial}^{\frac{1}{2}(s_x h_x+3)} \right]} = \Tr_{\partial} {\left[ \bigotimes_x \F_{x,\partial}^{\frac{1}{2}(s_x h_x+3)} \right]} = \prod_x d_x^{\frac{1}{2}(s_x h_x+3)}.
\end{equation}
	This is the first ingredient in arriving at an evaluated expression for eq. (\ref{eq:17})
	
	 The second trace, however, is somewhat more involved as a result of the terms coming from the symmetric subspace projector. To make this more manageable, we will break apart the trace as a trace over individual edges in the tensor network. To do so, we first need to look at the projector we have introduced. This operator consists of unitaries which are independently applied to the virtual Hilbert spaces surrounding a particular vertex. As such, we can split the tensor product of all such operators into operators acting upon both copies of a given virtual Hilbert space. Thus,
\begin{multline}
	\prod_x \Psym_x^{\otimes 2} \prod_{x\text{ s.t. }s_x=-1}\F_{x,v}^{\otimes 4} = \\
	\frac{1}{|G|^{2N}} \sum_{g_x,g_x',g_y,g_y'} \prod_{x \sim y}\left[\left((U(g_x)\otimes U(g_x'))\F_{xx'}^{\frac{1}{2}(s_x+3)}\right)\otimes \left((U(g_y) \otimes U(g_y'))\F_{yy'}^{\frac{1}{2}(s_y+3)}\right)\right].
\end{multline}
	Essentially, what we have done is written our product of projectors and conditional swaps as a product of operators acting upon Hilbert spaces associated with a particular edge in the tensor network and its copy (the primed variables). This includes the unitary representation on each virtual Hilbert spaces as well as the swaps applied. We have rewritten the sum as a sum over four independent group elements chosen from $G$ to specify that we have all terms where each virtual subspace in an edge and its copy may have a different group element attached to it. The $N$ in the prefactor is simply the number of vertices in our tensor network, coming from a subspace projector being applied to each vertex.

	We can write this expression a little more compactly in terms of the trace we wish to compute, leaving
\begin{equation} \label{eq:28}
	\Tr_{v}{\left[\bigotimes_{x \sim y}\ket{xy}\bra{xy}^{\otimes 2}\prod_x \Psym_x^{\otimes 2} \prod_{x\text{ s.t. }s_x=-1}\F_{x,v}^{\otimes 4}\right]} = \frac{1}{|G|^{2N}} \sum_{g_x,g_x',g_y,g_y'} \prod_{x \sim y} \Tr_{x \sim y}\left[ \Omega\right]
\end{equation}
	where the argument of the trace we wish to compute is
\begin{equation}
	\Omega = \ket{xy}\bra{xy}^{\otimes 2}(U(g_x)\otimes U(g_{x'}) \otimes U(g_y) \otimes U(g_{y'}))(\F_{xx'}^{\frac{1}{2}(s_x+3)}\otimes \F_{yy'}^{\frac{1}{2}(s_y+3)}).
\end{equation}
	The conditional swap operators here work in the same way as before, either flipping the spaces connected by maximally entangled states or not. We can thus combine this into one swap operator acting on either the $xx'$ subspaces or the $yy'$ subspaces conditioned by $s_x$ and $s_y$,
\begin{equation} \label{eq:39}
	\Omega = \ket{xy}\bra{xy}^{\otimes 2}(U(g_x)\otimes U(g_{x'})\otimes U(g_y)\otimes U(g_{y'}))(\F_{xx'}^{\frac{1}{2}(s_x s_y+3)} \otimes \I).
\end{equation}
	This then leads us to two possibilities. If no swap is applied, then maximally entangled states reside in the $x$ and $y$ virtual Hilbert spaces and $x'$ and $y'$ Hilbert spaces separately. Thus, our expression inside the trace becomes a tensor product of group elements $U(g_x g_y^{-1})$ and $U(g_{x'} g_{y'}^{-1})$ acting on disjoint Hilbert spaces after tracing out one virtual Hilbert space and its copy. The inverse comes from moving the operator from one Hilbert space to another through the maximally entangled state and multiplying group elements together.

	On the other hand, applying the swap connects maximally entangled states from $x$ to $y'$ and $y$ to $x'$. Now, when we take a partial trace we are left with just one operator acting on a virtual Hilbert space and its copy but not independently. This operator is given by $U(g_x g_{y'}^{-1} g_{x'} g_y^{-1})$. We can consolidate these two possibilities in the follow equation,
\begin{equation}
	\Tr_{x \sim y}{[\Omega]} =
\begin{cases}
	\Tr_{xx'}{[U(g_x g_y^{-1}) \otimes U(g_{x'} g_{y'}^{-1})]} = D_{xy}^2 \delta_{xy} \delta_{x'y'} & s_x=s_y \\
	\Tr_{xx'}{[U(g_x g_{y'}^{-1} g_{x'} g_y^{-1})]} = D_{xy} \delta_{xy'x'y} & s_x \neq s_y
\end{cases}
\end{equation}
	The traces are evaluated as Kronecker deltas because we have a trace of a regular representation (or direct sum of regular representations if the bond dimension is to be larger) such that the trace over any non-identity element is zero. This introduces a subtlety into the available states since the trace for $s_x \neq s_y$ can be nonzero for some nontrivial group elements residing on the lattice. Note that we still obtain the same factors $D_{xy}$ or $D_{xy}^2$ that appear in the trivial case following from a trace over swapped and non-swapped spaces.

	Returning to the original expression, we now have a product of delta functions dependent on our Ising variables which is being summed over group elements corresponding to each vertex in the tensor network. In order to fully evaluate the expression, we then need to count the number of terms in the sum over group elements for which the product of delta functions depending on the Ising variables is nonzero. 

	In order to perform this counting, we first partition configurations $\{s_x\}$ into subsets given by the number of domain walls, $\eta$, in the lattice. This will enable us to determine the number of factors in the product. 
	
	Let's first consider if we have one cluster (all $s_x=1$ or all $s_x=-1$) then no swaps are applied anywhere and we only have conditions of type $\delta_{xy}\delta_{x'y'}$. These conditions tell us that vertices connected by edges on the tensor network must have the same group element associated with it, $g_x = g_y$ for a given link. Thus, the only such terms which are nonzero in the sum are the ones for which all vertices in the tensor network have the same group element applied to each virtual index. This is independently and similarly true for the second copy of the tensor network. Thus, we have $|G|^2$ total nonzero terms in the sum since we have $|G|$ possible group elements residing on each vertex in the tensor network as well as its copy.

	Now, suppose we add one domain wall to the Ising lattice describing out tensor network so that we have two clusters. This changes terms associated with edge crossing the domain wall to four-index deltas, $\delta_{xy'x'y}$. These delta functions are associated with links connecting two regions in which the lattice is otherwise homogenous in group elements, i.e. crossing a domain wall. Each of these $\delta$'s associated with a particular domain wall describe the same condition, $g_{x} g_{y'}^{-1} = g_{x'} g_{y}^{-1}$. It is important to note that this condition is the same for each edge crossing the domain wall since group elements outside and inside the are the same for any nonzero term (following the $\delta_{xy}\delta_{x'y'}$ condition). Unlike the other type of factors in the product, this one only restricts one of the four groups elements associated with a link rather than two. This then gives us the freedom to determine one more set of group elements within the tensor network that would give nonzero terms, adding a multiplicative factor of $|G|$ to these terms. Thus, if we have $\eta$ domain walls, then there are $|G|^2|G|^{\eta} = |G|^{2+\eta}$ nonzero terms in the sum. So, returning to eq. (\ref{eq:28}),
\begin{equation}
	\Tr_v{\left[\bigotimes_{x \sim y}\ket{xy}\bra{xy}^{\otimes 2} \prod_x \Psym_x^{\otimes 2} \prod_{x\text{ s.t. }s_x=-1}\F_{x,v}^{\otimes 4}\right]} = |G|^{2+\eta-2N} \prod_{<xy>} D_{xy}^{\frac{1}{2}(s_x s_y +3)}.
\end{equation}
	Now, combining this expression with eq. (\ref{eq:14}), we obtain
\begin{equation}
	e^{-\A[\{s_x\}]} = \frac{|G|^{2+\eta-2N}}{\prod_x(D_x^2+D_x)} \prod_x d_x^{\frac{1}{2}(s_x h_x+3)}\prod_{x \sim y}D_{xy}^{\frac{1}{2}(s_x s_y+3)}, 
\end{equation}
	which gives us the associated Ising action
\begin{multline}
	\A[\{s_x\}] = -\sum_x \frac{1}{2}\log{d_x}(3+s_x h_x) - \sum_{x \sim y}\frac{1}{2}\log{D_{xy}}(3+s_x s_y) \\
	+\sum_x \log{(D_x^2+D_x)} - (2+\eta-2N) \log{|G|}.
\end{multline}
	We can write this in a slightly more useful form in which we can collect several terms as a constant that will be the same in both free energies associated with $Z_1$ and $Z_0$,
\begin{equation} \label{eq:36}	
	\A[\{s_x\}] = -\sum_x \frac{1}{2}\log{d_x}(s_x h_x-1) - \sum_{<xy>}\frac{1}{2}\log{D_{xy}}(s_x s_y-1) - \eta \log{|G|} + \text{const.}
\end{equation}

	We get the same expression for $Z_0$ but with a $h_x = 1$ everywhere. 
\end{proof}

	This is the Ising action for our locally symmetric random tensor network and forms the basis of a modified Ising model which lends itself to computing the R\'{e}nyi entropy which we do below.

\subsection*{Entanglement Entropy with Nontrivial Topological Term}
    
We begin computation of the entropy by considering a connected boundary region $A$ with reduced density matrix $\rho_A$. We will further allow the region $A$ to be disjoint which will be important due to the presence of the non-local term. The second R\'{e}nyi entropy $S_2(A)$ is given by Eq. (\ref{eq:8}). We can consider the numerator and denominator separately and denote them by $Z_1$ and $Z_0$, aligning with Lemma 1. The entropy average can be expanded in powers of fluctuations $\delta Z_1 = Z_1 - \overline{Z_1}$ and $\delta Z_0 = Z_0 - \overline{Z_0}$
\begin{equation} \label{eq:31}
	\overline{S_2(A)} = -\overline{\log{\frac{\overline{Z_1}+\delta Z_1}{\overline{Z_0}+\delta Z_0}}} = - \log{\frac{\overline{Z_1}}{\overline{Z_0}}} + \sum_{n=1}^{\infty} \frac{(-1)^{n-1}}{n}\left( \frac{\overline{\delta Z_0^n}}{\overline{Z_0}^n} - \frac{\overline{\delta Z_1^n}}{\overline{Z_1}^n} \right).
\end{equation}
 which has suppressed fluctuations at large enough bond dimension \cite{Hayden1}. Note that if we follow the previous result, this suppression is dependent on the physical bond dimension scaling with the virtual bond dimension and thus exceeding a physically relevant regime of small physical bond dimension.  We address this regime in the following subsection as this claim to suppressed bond dimension is necessary to demonstrate an area law expression. In any case, if the conditions are met for the fluctuations to be suppressed, then with high probability we can approximate the entropy with the averages performed separately over $Z_0$ and $Z_1$,
\begin{equation} \label{eq:26}
	S_2(A) \simeq - \log{\frac{\overline{Z_1}}{\overline{Z_0}}}.
\end{equation}
The '$\simeq$' is employed to describe an asymptotic equality that holds under appropriate conditions for which the fluctuation vanishes. This is the case for a bond dimension scaling exponentially in the volume of the region, and perhaps even for polynomially under certain physical assumptions \cite{Hayden1}. In any case, from Lemma 1 we see that $Z_1$ and $Z_0$ are both partition functions of classical Ising models so that the entropy is then a difference of free energies of these models,
\begin{equation}
	S_2(A) \simeq F_1 - F_0,
\end{equation}	
at an inverse temperature which is function of the bond dimensions. At large bond dimension (low temperature) the system will preference the ground state. In these circumstances, then, this difference of free energies can be approximated by a difference in ground state energies. In the $Z_0$ case, this ground state is a ferromagnet with all spins aligned in which the Ising action (Eq. \ref{eq:36}) is a constant. Note in particular that $\eta = 0$ in this case since the state is uniform in spin and there are no domain walls. 

In the $Z_1$ case, the ground state is a bit more subtle. If we allow the physical bond dimension to take on a large enough value as with the virtual bond dimension ($d_x \approx D_{xy}$), the ground state is characterized by a flip of spins in the region $A$ from the ferromagnetic field, introducing a $\log{D_{xy}} |\partial A|$ energy cost due to $|\partial A|$ non-aligned edges as well as an additional constant $\log{|G|}$ coming from the singular domain wall. Note further that if $A$ consists of $m$ disjoint regions, then the added constant becomes $m \log{|G|}$ now that the state possess $m$ domain walls. 

We can say then that, with high probability, the 2nd R\'{e}nyi entropy of the region $A$ scales asymptotically in bond dimension (inverse temperature) with the boundary of this region,
\begin{equation}
	S_2(A) \simeq - \log{\frac{\overline{Z_1}}{\overline{Z_0}}} = |\partial A|\log{D_{xy}}  - m \log{|G|} + o(1)
\end{equation}
where $\alpha$ is a constant dependent on the virtual bond dimension. Note that if our bond dimensions $D_{xy}$ and $d_x$ are equivalent and allowed to be arbitrarily large, then we have an area law entropy regardless of the shape of the region $A$. Moreover, we can see then how the topological entanglement entropy manifests in the entropy average as a difference in domain walls in the classical model ground states.

We can now attempt to present an expression for the Von Neumann entropy by way of an upper and lower bound. The Von Neumann entropy is lower bounded by the second R\'{e}nyi entropy so that it follows immediately from above that
\begin{equation} \label{eq:35}
	S(A) \geq |\partial A|\log{D_{xy}}-log{|G|} + o(1).
\end{equation}

Next, we wish to upper bound the Von Neumann entropy. We do this by first bounding the maximal Schmidt rank. Consider a state $\ket{\psi}$ described by an arbitrary $G$-injective PEPS tensor network. We then consider a bipartition of the physical Hilbert space into regions $A$ and $\overline{A}$. We can then act on the region $\overline{A}$ by a linear operator $M_{\overline{A}}$ composed of local operations which purifies the degrees of freedom in $\overline{A}$. This produces a state $\ket{\phi} = (M_{\overline{A}} \otimes \I) \ket{\psi}$ whose Schmidt rank is greater than or equal to that of $\psi$ state since $M_{\overline{A}}$ is acting only on region $\overline{A}$ by local operations and can therefore only increase the maximal Schmidt rank. However, since the state is $G$-injective, we have an invariance given by this bipartition, $|G|^{-1} \sum_{g \in G} U(g)^{\otimes |\partial A|} \ket{\phi} = P_{\partial A}(G)\ket{\phi} = \ket{\phi}$. For clarity, recall that $P(G)$ is our symmetric subspace projects, not applied along boundary edges. Our maximal Schmidt rank of a state $\rho_A = \ket{\psi}\bra{\psi}$ is then upper bounded by the rank of this projector since it is otherwise purified. This leads us to a rank of $|\partial A|$-many $D_{xy}$ subspaces, 
\begin{equation}
	S(A) \leq \max{(\log{(\rank{\rho_A})})} \leq \log{(\rank{P_{\partial A}(G)\ket{\phi}})} = \log{(D_{xy}^{|\partial A|}/|G|)} = |\partial A|\log{D_{xy}} - \log{|G|}.
\end{equation}
Thus, we have shown that our Von Neumann entropy is upper bounded by a general area law form in the G-symmetric case and is similarly lower bounded, though asymptotically so. Our Von Neumann entropy then asymptotically has the desired tight bound,
\begin{equation}
	S(A) \simeq |\partial A|\log{D_{xy}} - \log{|G|}
\end{equation}
where $\alpha$ is a constant dependent on bond dimensions, and $|\partial A|$ is the size of the boundary of region A. It is important to note that this requires the assumptions of large bond dimension under which which we can specify a R\'{e}nyi entropy of the appropriate form for a lower bound.
	
	
\subsection*{Assumption of Small Physical Bond Dimension}
	The previous result is relevant in regimes where the physical bond dimensions can be taken as arbitrarily large, such as the holography setting in which the topologically trivial result has been employed \cite{Hayden1}. There are two issues in the above logic, however, when we consider the regime of small physical bond dimension (allowing the virtual dimension to be large as before).
	
	The first issue is that we need to still be able to say that the $F_1$ free energy at low temperature is that of flipped spins in the region $A$ in order for the difference in free energies to give the appropriate area law expression. This is clearly not the case when $D_{xy} \gg d_x$ and for small regions $A$ where the cost of non-aligning spins will be overshadowing by the cost from any non-aligned edges. We can conceivably resolve this issue by further constraining the region $A$ to be such that it's surface area-to-volume ratio is sufficiently small. In other words, constrained such that $V\log{d_x} > |\partial A| \log{D_{xy}}$, or $|\partial A|/V > \log{d_x}/\log{D_{xy}}$. Another solution this issue may be to think of the region $A$ on a renormalized lattice in which the number of vertices in the region $A$ is growing more rapidly than edges on the region's boundary with each step in renormalization. In both of these instances, the goal is to create a situation in which is it preferential in energy for the spins in region $A$ to be flipped as any cost saved by reduction in the surface area would be outweighed by the cost of flipped spins in the volume of $A$. These conditions seem plausible and may be more easily demonstrated in their extremes, but are unable to present a rigorous argument here.
	
	The second issue is that we can't bring this result to a regime of small physical bond dimension without first accounting for the suppression of fluctuations in \ref{eq:39} in this same regime. A rigorous proof of suppressed fluctuations following the logic of \cite{Hayden1} requires the expression
	$\frac{\overline{(Z_1^{(n)})^2}}{(Z_1^{(n),\infty})^2}-1$ to be sufficiently small. Here, the expression uses higher order R\'{e}nyi entropy (described in Appendix A). The superscript $\infty$ indicates a regime of large bond dimension, i.e. the ground state per the correspondence. A connection made in the previous result is that this second moment, the average of $(Z_1^{(n)})^2$, can be computed by interpreting it as a partition function of a so-called $\text{Sym}_{2n}$-spin model. Thus, the above expression being sufficiently small is equivalent to the partition function of this $\text{Sym}_{2n}$-spin model at some bond dimension $D=d_x=D_{xy}$ remaining close in value to it's large $D$ limit. This can be shown to be the case for certain values of $D$ which givens a condition under which the fluctuations are suppressed with high probability. In the previous results, this is demonstrated for exponentially scaling bond dimension which applied to both physical and virtual bond dimensions. 
	
	So what do we do then when the physical bond dimension is kept small? It seems possible that we may make a similar argument by allowing the virtual bond dimension to be sufficiently large. In this case, we run into the first issue yet for a slightly different model, the $\text{Sym}_{2n}$-spin model. Assuming we've resolved the first issue, then, we might say that the appropriate constraints on region $A$ are such that the large $D_{xy}$ partition function of the $\text{Sym}_{2n}$-spin model is that same as when we allowed $d_x$ to grow large with $D_{xy}$, i.e. the have the same low energy configuration. Again, a more rigorous proof of such is challenging. From here, it may be possible to set a conservative bound by considering the minimum cost incurred as was done previously or even attempt a low energy expansion of the partition function. This is yet another challenging task given the assumptions that would need to be placed on the region $A$. Nonetheless, it seems possible that under these assumptions, fluctuations might be suppressed given the \textit{virtual} bond dimension scales exponentially, or perhaps even polynomially.
	
	We remark that while the correspondence we've constructed applies regardless of bond dimension, a rigorous account of the entropy area law with correction, (averaged of random tensor network states) only applies when both the physical and virtual bond dimensions are allowed to be arbitrarily large. It is thus left to the reader as an open question, albeit a conceivable one, whether or not the later statement may still be true for small physical bond dimension.

\section{Conclusions and Future Work} \label{sec:8}

In this paper we extended the method of \cite{Hayden1}, providing a connection between the computation of R\'{e}nyi entropy with a difference of free energies of an Ising model. We generalize this method to include a local symmetry at each tensor in the tensor network lending itself to a non-local term in the Ising model and correspondingly a topological entanglement entropy present in the area law. We argue that this form holds exactly at zero temperature as well as low temperatures where bond dimensions scale exponentially, perhaps polynomially, following previous results (see \cite{Hayden1}). While a specific case is presented, we expect these results generalize to any dimension and for any symmetry given by any local, finite group. We can also see how the topology of region $A$ presents itself in the topological term, adding a power to $|G|$ in the logarithm by enforcing more domain walls in the non-trivial free energy expression. Thus, we have shown that a random PEPS tensor network subject to a local symmetry has an area law with high probability at sufficiently large bond dimension (low temperature). This allows us to go beyond consideration of renormalization fixed point PEPS when demonstrating an area law with topological entanglement entropy by showing such an expression for generic tensor networks with local symmetry. 
	
The primary drawback of our approach is the difficulty in rigorously demonstrating the area law expression for generic random PEPS states in the regime of small physical bond dimension. While we suggest some possible assumptions under which we might recover the desired result in this regime, a more rigorous resolution t this drawback is left as an open question. A resolution to this question would be an important consideration if one is interested in extended these results into a condensed matter regime. Nonetheless, our results are still relevant in the large bond dimension regime which can be considered in holography as was done previously. Here, one can see how certain wormhole geometries (domain walls in the bulk) may give rise to a constant correction when the entropy is measured on the boundary. Following this direction, one might extend our results by considering states with different representations for each link as a connection to holographic coherent states as described in \cite{qi2017holortn1}. 
	
It's worth noting that our construction, alongside the original results from \cite{Hayden1}, provide a methodology for relating features of tensor networks to terms in a classical spin model. This might allow someone to extend our model to a larger class by including other features. For example, one may encompass all string-net models by way of generalizing the symmetry to include MPOs symmetries \cite{csahinouglu2014characterizing}. This would require determination of a symmetric subspace projector for any MPO symmetry or another way to represent a generalized symmetry on an otherwise random tensor network. Another avenue of interest may be to extend the G-symmetric random tensor network construction to one where additional tensors without associated physical Hilbert spaces are added to the network. It also appears straightforward to extend the correspondence to constructions like Projected Entangled Simplex States whereby our entanglement occurs on simplices constructed from multiple vertices rather than between adjacent vertices, perhaps giving way to classical spin models with several-body interactions. 
\vspace{0.5 cm}



\section*{Acknowledgements}
		
This work is part of IQIM, which is a National Science Foundation (NSF) Physics Frontiers Center (NSF Grant PHY-1733907),



\bibliographystyle{plain}
\bibliography{refs}



\appendix

\section{Higher R\'{e}nyi Entropies} \label{sec:4}
	In this section, we will extend our results to a computing higher R\'{e}nyi entropies. This is primarily done to demonstrate how the topological correction manifests in higher entropy expressions.  

	We generalize the second R\'{e}nyi entropy to the $n$-th R\'{e}nyi entropy defined by
\begin{equation}
	S_n(A) = \frac{1}{1-n}\log{\frac{\Tr{\rho_A^n}}{(\Tr{\rho})^n}}.
\end{equation}

	We choose variables $Z_1^{(n)}$ and $Z_0^{(n)}$ for our numerator and denominator, respectively. 
	
\subsection*{Classical Model Correspondence for Higher R\'{e}nyi Entropies}
	
	We can write a generalization of Theorem 1, extending the correspondence of these traces to partition functions of classical spin models. The proof for this will follow the same structure as before but will make use of some examples to elaborate as needed.

\begin{thm}[Projected N-copy Trace/Modified N-spin Ising Model Correspondence]
	Let $Z_1^{(n)} = \Tr{\rho_A^n}$ for a reduced density matrix on a region $A$ which is described by a PEPS with local invariance given by a finite group, $G$ (eq. \ref{eq:29}). Then the Haar average of $Z_1^{(n)}$ is equivalently described by the partition function of a modified classical $n$-spin model of spins residing on the same lattice with Ising action given by
\begin{equation} \label{eq:49}
	\A[\{\Gamma_x\}] = - \sum_x \log{d_x} \chi(\Gamma_x^{-1} h_x) - \sum_{<xy>} \log{D_{xy}}\left(\chi(\Gamma_x^{-1} \Gamma_y) - n \right) + \sum_{\eta_i}(\chi(\Gamma_{\eta_i})-n) \log{|G|} + \text{const}
\end{equation}
	where $d_x$ is the physical bond dimension of the tensor at $x$ and $D_{xy}$ the virtual space bond dimension between tensors at $x$ and $y$. The function $\chi$ counts the number of cycles of a permutation group element. $\Gamma_x$ is the permutation group element corresponding to the value of the spin on the same vertex. The variable $h_x$ is a boundary field which is an identity on the $n$ copies, $\I_x$, for $x \in \overline{A}$ and a cyclic permutation on the $n$ copies, $\C_x^{(n)}$, for $x \in A$. The non-local term is characterized by a permutation group element associated to a particular domain wall, $\eta_i$, between clusters, $\Gamma_{\eta_i}$, and the cardinality of $G$.

	If we instead have an average over $Z_0^{(n)} = (\Tr{\rho_A})^n$, then we have the same correspondence but without a boundary pinning field, i.e. $h_x = \I_x, \forall x$. 
\end{thm}

\begin{proof}
	We first consider $Z_1^{(n)}$ which can be modified by a generalization of the "swap trick" where we instead permute the $n$ copies of the state \textit{cyclically} in the region $A$, the action of which is given by the operator $C_A^{(n)}$. It's important to remember that the action of this variable is cyclic which will be relevant later in this section. With this in mind, we can write
\begin{equation}
	Z_1^{(n)} = \Tr{\rho_A^n} = \Tr{[\rho^{\otimes n} C_A^{(n)}]},
\end{equation}

	Our average on $Z_1$ is now an average over $n$ copies of $\ket{V_x}\bra{V_x}$. This average is a symmetric subspace projector of the $n$-fold tensor power Hilbert space \cite{Harrow1} so that
	
\begin{equation}
	\overline{\ket{V_x}\bra{V_x}^{\otimes n}} = \frac{1}{C_{n,x}} \sum_{\Gamma_x \in Sym_n} \Gamma_x
\end{equation}
	where we replace the $g_x$ notation from previous work with $\Gamma_x$ to specify group elements coming from the $n$-copy average and not elements of our local symmetries groups. This is to avoid confusion between the two and keep consistency with our former definition for $g_x$. These $\Gamma_x$ are permutation group elements acting upon $n$ copies of a single site Hilbert space but can be non-cyclic group elements unlike the single site operation by $C_A^{(n)}$. The normalization constant 
\begin{equation} \label{eq:52}
	C_{n,x} = \frac{(D_x+n-1)!}{(D_x-1)!}
\end{equation}
	remains the same as in previous work, representing a sum of traces over each group element in $Sym_n$ acting upon each single site Hilbert space of vertex $x$.

	From here, we can split the trace into virtual and physical Hilbert space traces as done in Section \ref{sec:3} since the permutations still act independently on these Hilbert spaces and the local symmetric projectors are still acting on the virtual Hilbert spaces. The physical trace then shows up in the Ising action in the same way as in \cite{Hayden1}. We will give the expression for this term when with the full expression for the Ising action below.

	Before we do this we must first deal with the trace over virtual Hilbert spaces. The permutation group elements act in a similar way to the swap operators from above. Rather than expressing them as operators depending on Ising variables, however, we leave the operators in the general from of $\Gamma_x$ to indicate that they are some particular permutation group element acting upon all copies of a particular virtual subspace of $x$. This allows us to rewrite equation \ref{eq:28} as
\begin{equation}
	\Tr_{v}{\left[\bigotimes_{<xy>}\ket{xy}\bra{xy}^{\otimes n}\prod_x \Psym_x^{\otimes n} \prod_{x}\Gamma_{x,v}^{\otimes 4}\right]} = \frac{1}{|G|^{nN}} \sum_{\substack{g_{x_1},\cdots,g_{x_n} \\ g_{y_1},\cdots,g_{y_n}}} \sum_{\{\Gamma_x\},\{\Gamma_y\}} \prod_{<xy>} \Tr_{<xy>}\left[ \Omega\right]	
\end{equation}
	where $\Gamma_{x,v}^{\otimes 4}$ represents the action of $\Gamma_x$ of the four $n$-copy virtual Hilbert spaces at vertex $x$. Note the additional sum which runs over all combinations of $\Gamma_x$ and $\Gamma_y$. Since we are now working with $n$-copies of our state, we have a local symmetry projector acting on each of the copies resulting in a prefactor of $\frac{1}{|G|^{nN}}$ and we have $n$ group elements associate with each copy for both sides of a particular edge on the tensor network. The $\Omega$ in the above expression is constructed in the same way and is given by
\begin{multline}
	\Omega = \ket{xy}\bra{xy}^{\otimes n} (U(g_{x_1}) \otimes \cdots \otimes U(g_{x_n}) \otimes U(g_{y_1}) \otimes  \cdots \otimes U(g_{y_n})) (\Gamma_{x,v} \otimes \Gamma_{y,v}).
\end{multline}

	Performing a permutation on the virtual Hilbert spaces copies of vertex $x$ and on vertex $y$ is the same as performing a different permutation on $x$ or $y$ and an identity on the other. This new permutation on $x$ is given by a composition of $\Gamma_{x,v}$ and $\Gamma_{y,v}^{-1}$, the inverse group element. Since this is still just an arbitrary group element we can just leave $\Gamma_{x,v}$ and make $\Gamma_{y,v}$ an identity, mirroring eq. \ref{eq:39}.

	Since composition of group elements is again a group element, we can reduce the sum over $\Gamma_x$ and $\Gamma_y$ to just a sum over all possibilites for $\Gamma_x$ where $\Omega$ becomes
\begin{equation}
	\Omega = \ket{xy}\bra{xy}^{\otimes n} (U(g_{x_1}) \otimes \cdots \otimes U(g_{x_n}) \otimes U(g_{y_1}) \otimes  \cdots \otimes U(g_{y_n})) (\Gamma_{x,v} \otimes \I).
\end{equation}
	
	We can now consider different cases for different permutation group elements, $\Gamma_{x,v}$. The action of this operator permutes the group elements in the virtual Hilbert space and its copies associated with one vertex of a given edge. We now consider the permutation of group elements on each side as a relative shift in the copies associated with vertex $x$ and $y$, respectively. This can be given by a single permutation of group elements on one side.  Since the maximally entangled states associated with each edge are attached to a particular copy of the edge Hilbert space, permutation of operators on these spaces then allows us to push the unitaries on one side of the edge through as inverses in every possible configuration.

	At this point we introduce what is called an edge group element, $e_i = g_{y_i^{-1}} g_{x_i}$, to make our expression simpler. It is easy to see why this is useful if the permutation is an identity since we're doing no more than pushing together each copy of the edge independently. If multiple such copies of an edge are dependent, then we get a unitary of a product of these edge group elements. For example, if the permutation is a flip on two copies of a Hilbert space and an identity on the rest then we can write some unitary operator $U(g_{x_i}g_{y_j}^{-1}g_{x_j}g_{y_i}^{-1})$ for some $i$ and $j$ tensored with a unitary over one edge element for every other copy. Since a trace is invariant under cyclic permutation of operators within the trace, it follows that the nontrivial unitary can also be written as $U(g_{y_i}^{-1}g_{x_i}^{-1}g_{y_j}^{-1}g_{x_j}) = U(e_i e_j)$. This can be extended to any number of dependent copies of an edge such that we can always write our unitaries in this way. Note that these unitaries are unique up to cyclic permutation of edge group elements.

	With this in mind, $\Gamma_{x,v}$ then connects copies of Hilbert spaces on one side of an edge to copies on the other side which gives rise to tensors of all possible unitaries over edge group elements where each edge group element shows up only once in the expression. This allows us to define a map from $\Tr_{<xy>}\left[\Omega\right]$ to $\Tr_{x}\left[\Omega'\right]$ given by translating the cycles associated with a particular permutation group element to unitaries over edge group elements corresponding to these cycles. Here, the trace over $x$ indicates a trace over the Hilbert space and its copies for one side of a particular edge (we've brought operators through maximally entangled states and reduced the other side to an identity). 
	As a specific example, suppose $\Gamma_{x,v}$ is given by the cycles $(23)(145)$ for the group $Sym_5$ where $(ijk)$ describes a cyclic permutation of operators in Hilbert spaces $i$, $j$, and $k$ (in this case we have ordered an edge Hilbert space and its copies in some way). Then the instance of $\Omega$ containing this particular permutation can also be written as $\Omega' = U(e_2 e_3) \otimes U(e_1 e_4 e_5)$ where each unitary operator acts on the Hilbert space associated with its particular edge group elements. Note that these unitary operators are unique up to cyclic permutations just as the indices in the cycles of a permutation group element are.

	Enumerating these possibilites for large order permutation groups is burdensome, but we present the cases for $Sym_4$ as an example. Here, $i,j,k,l \in \{1,2,3,4\}$ with none of these indices being equal.
\begin{equation}
	\Tr_{<xy>}{[\Omega]} =
\begin{cases}
	\Tr_{x}{[U(e_1) \otimes U(e_2) \otimes U(e_3) \otimes U(e_4)]} = D_{xy}^4 \delta_1 \delta_2\delta_3\delta_4 & \Gamma \leftrightarrow (1)(2)(3)(4) \\
	\Tr_{x}{[U(e_ie_j) \otimes U(e_k) \otimes U(e_l)]} = D_{xy}^3 \delta_{ij}\delta_k \delta_l & \Gamma \leftrightarrow (ij)(k)(l) \\
	\Tr_{x}{[U(e_ie_j) \otimes U(e_ke_l)]} = D_{xy}^2 \delta_{ij}\delta_{kl} & \Gamma \leftrightarrow (ij)(kl) \\
	\Tr_{x}{[U(e_ie_je_k) \otimes U(e_l)]} = D_{xy}^2 \delta_{ijk} \delta_l & \Gamma \leftrightarrow (ijk)(l) \\
	\Tr_{x}{[U(e_ie_je_ke_l)]} = D_{xy} \delta_{ijkl} & \Gamma \leftrightarrow (ijkl)
\end{cases}
\end{equation}

	We see that considering the trace this way again gives us a product of Kronecker deltas as before. Returning to the connection to an Ising model, we have the same $Sym_{n}$-spin model with the same boundary field as in (Hayden et. al.) but we now obtain additional factors of $|G|$ depending on solutions to equations given by the product of delta functions. It is important to note that the expression is otherwise identical to the previously obtained result. The higher order entropy permutation affects the number of nonzero terms in a sum of expressions that are otherwise unchanged by the topological symmetry. 

	There is a subtle difference, however, in that the number of additional solutions across each domain wall in the $Sym_n$-spin model is not simply $|G|$ but a multiple thereof. Nontrivial unitaries in the trace (those containing more than one edge group element) give rise to Kronecker deltas with additional solutions. Each nontrivial Kronecker delta gives another set of $|G|$ solutions so that the additional factors of $|G|$ introduced at each domain wall depends on the relative permutation across the domain wall. We quantify the number of these extra solutions by using $\Chi(\Gamma_{\eta_i})$ to indicate the number of cycles in the relative permutation across a domain wall indexed by $\eta_i$.

	The number of added solutions at a given domain wall is then $n-\Chi(\Gamma_{\eta_i})$. This coincides with no additional solutions if there is not a domain wall ($\Chi(\I)=n$), one additional solution if the relative permutation is a swap on two copies ($\Chi(\F_{ij} \otimes \I)=n-1$), and $n-1$ additional solutions if the relative permutation is a cyclic permutation ($\Chi( (1 \cdots n ) ) = 1$). The prefactor can then be written in a slightly more complicated form as before so that
\begin{equation}
	\Tr_v \left[ \bigotimes_{<xy>} \ket{xy}\bra{xy}^{\otimes n} \prod_x \Psym_x^{\otimes n} \Gamma_{x,v}^{\otimes 4} \right] = |G|^{n(1-N)} \prod_{\{ \eta_i \} } |G|^{\Chi(\Gamma_{\eta_i})} \prod_{<xy>} D_{xy}^{\Chi(\Gamma_x \Gamma_y^{-1}) - n}.
\end{equation}
	The product over $\eta_i$ is a product over all domain walls in a particular configuration, $\{\Gamma_x\}$, of the Ising model lattice. Notice how the separate product over bond dimensions translates to the same result as previous work, but we now have a power of $|G|$ translating to the topological entanglement entropy as we will we see more clearly below.
	
	In a more concise way, we are considering a particular configuration, determining the relative permutation across a domain wall, and then constructing an equivalence to a system of equations given by a product of Kronecker deltas. We can systematically count the number of solutions to determine the addition factors of $|G|$. From here we can write the Ising action as
\begin{equation}
\begin{split}
	\A^{(n)}[\{\Gamma_x\}] = 
	& - \sum_{<xy>}\log{D_{xy}}\left( \Chi(\Gamma_x^{-1}\Gamma_y) - n \right)  - \sum_{x} \log{d_x} \Chi(\Gamma_x^{-1} h_x) \\
	& + \left(nN - n + \sum_{\eta_i} (\Chi(\Gamma_{\eta_i}) - n)\right) \log{|G|}  + \sum_x \log{\C_{n,x}}.
\end{split}
\end{equation}
	with the boundary field, $h_x$, given by
\begin{equation}
h_x = 
\begin{cases}
	\I_x, & x \in \bar{A} \\ 
	\C_x^{(n)}, & x \in A.
\end{cases}	
\end{equation}
	Grouping the constant terms in the Ising action, we recover eq. \ref{eq:49}. 
	
	For $Z_0^{(n)}$, the same analysis holds with a trivial boundary field whereby $h_x$ is an identity everyhwere.
\end{proof}
	
\subsection*{Higher R\'{e}nyi Entropies for a Nontrivial Topology}	
	Now, we attempt to write down the form of a generalized R\'{e}nyi entropy of a random PEPS subject to a non-trivial topology. Applying Theorem 2, we can again write the entropy as a difference of free energies,
\begin{equation}
	S_n(A) \simeq \frac{1}{1-n} \log{\frac{\overline{Z_1^{(n)}}}{\overline{Z_0^{(n)}}}} \simeq \frac{1}{1-n}(F_1^{(n)} - F_0^{(n)})	
\end{equation}
	where the superscript on the free energies means that we are referring to the free energy of our $n$-spin model following the correspondence. At large bond dimension, or low temperature, the free energies are approximated by ground state energies. To determine the ground state, we note that the internal bond dimension term in the Ising action is zero when $\Gamma_x = \Gamma_y$, encouraging a ferromagnetic state in which the $2n$-spins are aligned. The boundary terms vanish when the particular cycle is aligned with the boundary magnetic field so that $\Gamma_x$ is a cyclic permutation within the magnetic field and $\I$ otherwise. This encourages a uniform identity state without a magnetic field and exactly $m$ domain walls around the region $A$ with a magnetic field. The relative permutation across these domain wall is a cyclic permutation so that the sum over domain walls is simply $1-n$. Thus,
\begin{equation}
	\overline{Z_1^{(n)}} \Big|_{D \rightarrow \inf} \simeq \text{const.} \times e^{(1-n)\log{D}|\partial A| - (1-n) m \log{|G|}}
\end{equation}
	Without the magnetic field, there is a ferromagnetic phase in which the ground state energy is a constant. So, we can write the difference of free energies and therefore the entropy as
\begin{equation}
	S_n(A)\Big|_{D \rightarrow \inf} = |\partial A| \log{D} - m \log{|G|} +o(1). 
\end{equation}
	This recovers the area law form as desired. We do this in the case where $d_x = D_{xy} = D$ such that both physical and virtual bond dimensions are allowed to be arbitrarily large. It remains as a secondary open question if the small physical bond dimension case can be accounted for with higher R\'{e}nyi entropies as for the second.

\end{document}